# Unveiled electric profiles within hydrogen bonds suggest DNA base pairs with similar bond strengths

Y. B. Ruiz-Blanco[1], Y. Almeida[2,#a,#b], C. M. Sotomayor-Torres[3,4,5] and Y. García[6,#c,*]

[1]Facultad de Química y Farmacia. Universidad Central "Marta Abreu" de Las Villas, Santa Clara, 54830, Villa Clara, Cuba.

[2]Systems Biology Direction, Center of Molecular Immunology, Havana, 11600, Cuba.

[3]Catalan Institute of Nanoscience and Nanotechnology (ICN2), CSIC, Campus UAB, Bellaterra, 08193 Barcelona, Spain.

[4]The BIST, Campus UAB, Bellaterra, 08193 Barcelona, Spain.

[5]Instituciò Catalana de Recerca i Estudis Avançats (ICREA), 08010 Barcelona, Spain.

[6]Independent Researcher, 08290 Barcelona, Spain.

#a Current Address: The Hamburg Centre for Ultrafast Imaging, 22607 Hamburg, Germany.

#b Current Address: Department of Chemistry, Institute for Biochemistry and Molecular Biology, University of Hamburg, Martin-Luther-King-Platz 6, 20146 Hamburg, Germany.

#c Current Address: Departament of Mathematics, Polytechnical University of Catalonia, ESAB, Esteve Terradas, 8, E-08860 Castelldefels, Spain.

[*]Corresponding author email: yamila.garcia@uv.es.

# Abstract

Electrical forces are the background of all the interactions occurring in biochemical systems. From here and by using a combination of *ab-initio* and *ad-hoc* models, we introduce the first description of electric field profiles with intrabond resolution to support a characterization of single bond forces attending to its electrical origin. This fundamental issue has eluded a physical description so far. Our method is applied to describe hydrogen bonds (HB) in DNA base pairs. Numerical results reveal that base pairs in DNA could be equivalent considering HB strength contributions, which challenges previous interpretations of thermodynamic properties of DNA based on the assumption that Adenine/Thymine pairs are weaker than Guanine/Cytosine pairs due to the sole difference in the number of HB. Thus, our methodology provides solid foundations to support the development of extended models intended to go deeper into the molecular mechanisms of DNA functioning.

# Introduction

The understanding of dynamic events at atomic level in DNA structures has become one of the most relevant goals in order to face frontier challenges in nanotechnology [1-5]. DNA stability results from a combination of biochemical processes and as such it is assisted by electrical forces. Understanding the relative strengths of these forces involved in the specific bonding of DNA would provide physical foundations for molecular understanding of biology [6,7]. Given DNA's double strand is stabilized by hydrogen bonds (HB) which hold the two chains together [1,6-13], we propose a methodology to describe the features of electric field (**E**) along HB and to estimate the relative strengths of the different types of HB that hold the strands of the double-helix structure in DNA.

DNA exists in a folded-double-helical arrangement [6]. The role played by HB forces in the stability of DNA base pairs, Adenine/Thymine (A/T) and Guanine/Cytosine (G/C) respectively, is widely discussed in literature [1,7,8]. The question about how important could be the contribution of individual HB has been already raised [11,13,14]. Replication, recombination, translation and even site directed modification of DNA sequences are processes mediated by HB and therefore it is a generalized view that the role of HB in such events is crucial [1,6,7,15,16]. However, current analyses regarding intrabond forces associated with HB connecting the base pairs in DNA do not show consensus [11,17]. Often, the different influence of A/T and G/C pairs is interpreted in terms of the number of HB [7,12,18]. This commonly accepted hypothesis is based on the assumption of similar strength contributions for all the HB. We highlight two problems associated with such simplified approach: (i) one related with the different electromechanical configuration for each HB pair and (ii) a subtler one associated with the interpretation of HB as electrical dipoles even in cases where near-field interactions are determinant. (Here and along this report, we will call models describing phenomena at interaction distances larger than atomic dimensions as far-field approaches, conversely those models representing the effects at distances similar to atoms size will be referred as near-field approaches).

Concerning the first point, the reliability of such interpretations based on equal strengths of the HB would rely on the assumption of identical atomic configurations [7,12]. Even when this assumption might work well to explain the average effect of HB in processes of far-field scope and where the relevance of atomic contact mechanisms becomes negligible, events such as zipping and unzipping of DNA base pairs, where HB are formed or broken one by one, definitely, do not satisfy this premise.

Regarding the second point, the use of a dipole model to describe HB interactions permit the interpretation of macroscopic thermodynamic events such as macromolecular dimerization occurring on far-field regimes. However, in processes controlled by near-field interactions, e.g protein docking or DNA translation, the dipole model should be place on doubt in favour of atomic-contact-resolved approaches. Upon these bases the electric dipole based description of HB should be assumed with caution [9,11].

Aligned to the previous reasoning and with the aim to contribute to current debates [7,12,13,18] we approach an answer to the central question: *to what extent the three HB of G/C pairs related to the only two of A/T pairs could suggest a larger thermodynamic stability?*

In order to contribute to address this issue, we introduce a model for A/T and G/C base pairs, and develop a method to numerically evaluate single HB strengths. The identification of a mean-field parameter quantifying relative HB strengths is as an initial element for further extended models that, while including a huge number of structures as well as proper analysis of statistical events, could help to address questions as such here raised. Our method is based on the development of accurate quantum calculations as well as continuous elastic models. Initially, we show that the electric features for each HB differ in both types of base pairs, thus becoming in an electrical signature of HB identity. Then, we evaluate the relative strengths among the selected HB by introducing a simplified elastic model. Further interpretations lead us to conclude that efforts on the understanding and control of molecular events in DNA, might benefit from considering the renewed interpretation of HB here introduced that pays attention to the inherent electrical nature of such interactions.

# Methods

## Model Structures Definition

AT (A 29 and T 15) and GC (G 30 and C 14) were selected from PDB 1W36 [19]. Hydrogen atoms were added using the tool AddH implemented in UCSF Chimera [20]. The selection of the structure obeys to the criteria of selecting initial model systems starting from physically relevant events. For the definition of the final structure, i.e. the one used for numerical calculations, atomistic optimizations in the relevant cluster are systematically developed [16,21].

## Numerical Modelling

The atomistic model comprises the use of improved nonlocal DFT to describe atomic systems [22-24]. Our approach is specialized to deal with varying electron density systems, as required for a first-principles based description of HB interactions. The approach comprises the solution of a quantum mechanical ensemble in vacuum. Thus, we first define a reduced atomic model, region indicated in (Figs 1a and 1c. Then, we introduce mean field DFT potentials to account for such effects derived from electronic interactions. Our improved DFT methodology accounts for dispersion forces in a systematic manner and provides self-consistent exchange-correlation potentials to solve intrinsic many-body problems while retaining the advantages of the mean-field approach. Finally, the nonlocal exchange-correlation potential used in this specific problem comprises adjustable nonlocal Fock exchange in addition to the local exchange in PW91 functional (see S1 File). Electronic states are described on the basis of 6-31g (d,p) basis sets considering the influence of polarization functions [25,26]. To quantify how the numerical methods influence the mean values here reported we introduce a water dimer. This reduced scenario facilities the understanding of the concepts as well as it settles a comprehensive frame for design and optimization of our methods (see S2 File).

Full version is available on: http://journals.plos.org/plosone/article?id=10.1371/journal.pone.0185638

# Computational Details

The code GAUSSIAN09 is used for numerical DFT calculations [26]. In S1 File we provide details on the numerical route we follow to implement our method. Such approach could be also followed by other quantum chemistry codes and it is an important advantage of the methodology we offer here. The convergence of our calculations is fixed on the variation in the target quantity by less than 10% when increasing the quality of any the numerical parameters. Numerical parameters here refer to bases sets and other self-consistency parameters intrinsic to the computational codes used.

Full version is available on: http://journals.plos.org/plosone/article?id=10.1371/journal.pone.0185638

# Results

## Going inside HB in a model DNA structure: *the electrical signature*

With the aim to introduce the analysis of electric field profiles within HB in DNA structures we develop the following procedure. To describe electrical features with intrabond resolution we use a methodology previously developed by some of us which accounts for non-dispersive forces while going beyond Density Functional Theory (DFT) as a frame to develop *ab-initio* calculations [22-24,27]. Then, we introduce a characterization in terms of averaged **E**. Consequently, we model HB in DNA attending to the spatial evolution of **E** within the region defined by the pair interacting atoms, see Fig 2. Henceforth, we carried out numerical calculations based on nonlocal DFT as well as using the advantages of Green Function methods [27]. We calculate stationary potential energy surfaces, which are modulated by moving a virtual point charge. By keeping record of such variations we estimate **E**=**E**(x,y,z) as the static limit of actually occurring dynamical events.

To demonstrate the feasibility of our approach when applied in a model DNA structure we use the coordinates of two model clusters associated with representative A/T and G/C base pairs. Such coordinates [19] are obtained from a crystallographic structure (PDB 1W36) and are subsequently optimized (see Methods section for further details). Reduced atomistic representations are sketched in Figs 1a and 1c. To assist on the interpretation of our numerical results we use a simple model system constituted by the water dimer (details in the S2 File). This reduced scenario for the definition of a HB, facilitates the understanding of the concepts and also sets a comprehensive frame for the design and optimization of our *ab-initio* techniques [27,28].

Fig.´s 3 and 4 show the analysis of **E** behaviour along the main axis for each HB in the base pairs, A/T and G/C respectively. Primarily, we notice very *high* values for |**E**| along the analysed space that also retains very *high* minimum values (~10 V nm$^{-1}$) in a transition region defined at convenience on the surroundings. Note that similar values are being indicated as a possible reference in terms of **E** required to perturb HB in DNA [29]. The finite **E** threshold value refers to the minimal value that must be exceeded to distinguish electrical responses according to the regime and effects analysed. Above this threshold the limits to maintain linear regime characterizations of electrical related events should be assumed with caution. Using a classical mechanics analogous this result can be interpreted as follows. The "threshold" concept makes reference to the minimal value required to activate a static response. The analogous parameter in a classical mechanic's context would be "the static friction coefficient". The existence of such electric threshold seems essential for further understanding on the electromechanical responses in DNA [29]. Notice, for instance, that HB in the DNA model have lower threshold value than the HB in the water dimer. This fact could be interpreted in terms of lower electrical inertia for HB in the DNA model when comparing with the behaviour of HB in the water dimer.

We would also like to remark that present experimental techniques could not yet provide estimations of the electrical forces inside a chemical bond and then we cannot proceed via a direct comparison for our indexes [30,31]. Nevertheless, we carried out an indirect comparison of our methodology, and for that purpose we analyse the water dimer in terms of (i) predicted structural geometries vs. experimental measurements (to test our computational methods) (ii) predicted single bond forces vs. accepted interpretations for the relative forces between covalent and HB in the water dimer (almost 10 times superior in strength respectively). Such elements are discussed in the first section of the S2 File.

Full version is available on: http://journals.plos.org/plosone/article?id=10.1371/journal.pone.0185638

From our point of view, the method here introduced contributes to complete current characterization techniques of directional chemical bonds, e.g. HB [9]. The description of HB through intrabond electrical parameters, e.g. threshold values, $t_a$ descriptors, could be an initial precursor for systematic quantification of the electrical nature of chemical bonds (see transition regions in Figs 3 and 4). In addition, the benefit of obtaining descriptors associated to the interactions in response to an electric field, should favour the design of nanoelectronic devices based on the selective electrostatic tuning of the most susceptible chemical bonds.

## HB contribution to DNA stabilization: *3 could be less than 2*

To approach the understanding on how electrical HB could influence the mechanical response of DNA representative base-pairs, we map the information obtained from previous analysis in a reduced elastic model. The different values obtained for $t_a$ in the characterization of HB, Figs 3c and 4c, are then interpreted from a mechanical point of view. Thus, we find out a relationship between $t_a$ and an effective elastic constant, $k^{eff}$. The choice of an elastic model to describe features on chemical bonds in the limit of small deviations from equilibrium positions is widely used [18]. Here, we introduce a method to estimate $k^{eff}$ of DNA's HB by mapping HB into 1D-springs of negligible mass connecting boxes at atom positions, see Figs 1b and 1d. By *ad-hoc* reasoning we introduce $k^{eff}$ describing such HB springs as an inverse function of the previously defined $t_a$ descriptor ($k^{eff} = k^{eff}[t_a^{-n}]$). The $t_a$-coefficient deals with the delocalization of the electron cloud. As the $t_a$-values increase, the electron density deviates from the bond symmetry and thus the interaction force weakens. (See also S2 File for a comparison between HB and covalent bonds (CB) in the reference water dimer, an argument in favour of this hypothesis). For instance, in the water dimer, the $t_a$-values for the CB are one order of magnitude below the $t_a$-value for the HB. It is also accepted that CB are approximately *ten* times stronger than HB. Under such assumptions we formulate the relationship between the $t_a$-parameters and the strength of the HB by doing $n=1$ in the expression for the definition of $k^{eff}$. The same relationship could be applied to other systems and therefore it should be possible to describe the strength of a single bond from static structural information, i.e. by using the $t_a$-parameter introduced in the present study derived from the atomic positions.

We then estimate the effective elastic constants for the five HB under analysis. Using the force constant of the water dimer as reference, see S2 File, we obtained the following relative ratios of the strengths of DNA's HB related to solvent (water)'s HB: $HB_1$=2.3, $HB_2$=1.9, $HB_3$=1.9, $HB_4$=0.9 and $HB_5$=0.8. Even when such numbers would suffer due to inaccuracies in the estimation of $n$, we could safely assure that our calculations bring noticeable differences in terms of strengths associated to HB stabilizing DNA, and the effective contribution of the two first HB (i.e. the ones representing an A/T) could be superior to the mechanical strength contribution of the other three HB (i.e. the ones representing a G/C). Then, we issue an indicative warning in regards to the fact that three HB in G/C base pair, do not *necessarily* provide a stronger interaction than the two HB in A/T pairs as was interpreted from some previous larger scale experimental results [7,12].

Therefore, here we have found that in electromechanical terms A/T pairs could be as strong as G/C pairs. These results can be influenced by environmental elements like solvent, ions, pH and adjacent inter-bases stacking interactions [30,32-34]. Also, further studies are needed in order to depict how these electromechanical properties are influenced by the presence of an external **E** induced by other DNA molecule, RNA or DNA binding proteins e.g. helicases, topoisomerases. Noticeably, the statements and the conclusions that are made here could be extended to other HB-systems out the scope of DNA.

Full version is available on: http://journals.plos.org/plosone/article?id=10.1371/journal.pone.0185638

# Discussion

We have shown that DNA base pairs are stabilized by the contributions of well distinguishable HB. Further numerical evaluation of the relative strengths associated to such single HB in model A/T and G/C base-pairs lead us to issue a warning regarding the widely accepted criteria that the superior thermodynamic stability of G/C pairs when compared with A/T pairs, is originated on the different number of HB stabilizing the structures. Our claim goes in favour of pushing for the motivating appealing of going into intrabond scales to appreciate and quantify the richness of well localized atomic events. This need of describing events from an atomistic perspective [35], is even today superseded by the systematic use of thermodynamic criteria which are no longer accepted when dealing with single atomic events. Therefore, in light of the results here presented and with the aim to shed light on the initial motivating question highlighted in this manuscript, we could safely state that the thermodynamical stability of DNA must be analysed attending to *well distinguishable individual HB contributions which are actually zipping (stabilizing) DNA*. In summary, our approach is based on the development of *ab-initio* methods to obtain numerical results regarding the electrical characteristics inside HB and the mapping of such results in a mean-field electrical parameter $t_a$ which is obtained after the introduction of an *ad-hoc* mechanical model. Two main advantages should be highlighted regarding our method. First, the description of the electrical interactions taking place within the bonding region and further mapping of such description in a single mean-field parameter. The handling of this feature might be beneficial to model critical phenomena like the electromechanical response of diverse hydrogen-bonded systems and might facilitate the understanding of still open life-essential events [7,15,16,36]. Second, present numerical methods suffer from technical uncertainties when describing HB due to the limited consideration of the electron-electron interactions in model systems, and such dispersive forces are one of the main contributors to HB formation and further stabilization [10,22]. Our methodology, within DFT formalisms, represents a flexible and feasible way of quantifying reduced systems. Results derived from our approach could be incorporated in extended schemes intended to approach large scale thermodynamic events.

Finally, our results could contribute to understand previous controversial results regarding the role of HB in the electromechanical response of DNA [15,17,29]. However, more work is needed in this respect since the exact functional relation between the structural $t_a$-coefficient and the individual HB strength might be difficult to estimate given the inability of an elastic model to capture the realm of a complex biological system as DNA base-pairs represent. Our message regarding HB relative strengths in DNA should be then interpreted as a major warning. Thus, our methodology should be handled in the context of a seminal development dealing with the foundations of established paradigms. From our point of view, such established criteria regarding HB contributions to DNA stability should be reviewed if we plan to deal with atomic events. Looking further ahead, we expect that our results would contribute to solve fundamental issues such as specific gene editing and manipulation [3,4,37,38] and to facilitate realizations of electromechanical DNA based devices [39-42] while surpassing the technological barriers associated to measurement of intrabond forces [13,17]. As new experimental tools are developed [30,31], we expect an essential role for theoretical methodologies in achieving the goal of true atomic-level control and helping to pave the way for the promising biomolecular engineering.

# Acknowledgments

CMST acknowledges the financial support from the Spanish MINECO project PHENTOM (FIS2015-70862-P), the programme Severo Ochoa (Grant SEV-2013-0295) and funding from the CERCA Programme/Generalitat de Catalunya. Computational facilities have been provided by "Centro de Supercomputación de Galicia" (CESGA).

Full version is available on: http://journals.plos.org/plosone/article?id=10.1371/journal.pone.0185638

# Supporting Information

**S1 File. On the design of our DFT method.**

**S2 File. Understanding electrical forces in the ubiquitous water dimer.**

**S3 File. Definition of a numerical descriptor for a quantification of the electrical forces in directional bonds.**

Full version is available on: http://journals.plos.org/plosone/article?id=10.1371/journal.pone.0185638

**Figures List**

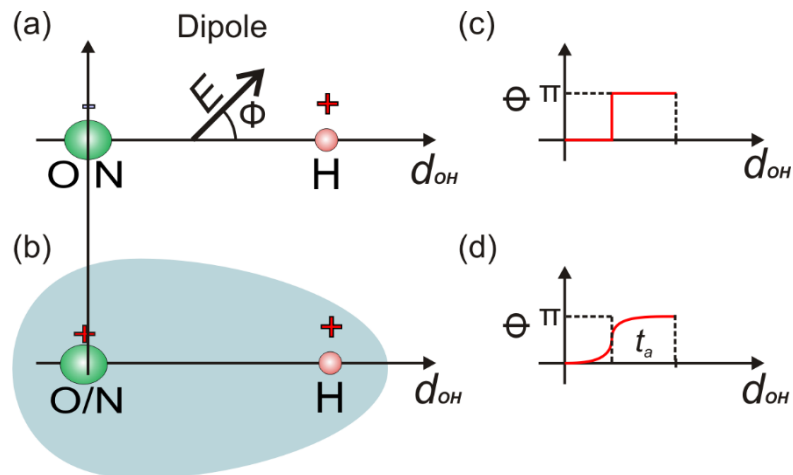

Fig 1. Schematic representation for HB connecting base pairs in DNA. **a, b**) A/T pair models. **c, d**) C/G pair models. On the left panels the atomistic structural representations are included. HB are labelled as $HB_1$, $HB_2$, $HB_3$, $HB_4$ and $HB_5$ throughout this work. On the right panels we include a diagram of the reduced elastic configuration we have used. A, T, G, and C basis are considered as rigid 2D boxes with identical mechanical responses. Then, the differences regarding molecular strengths will be originated on the 1D springs modelling HB.

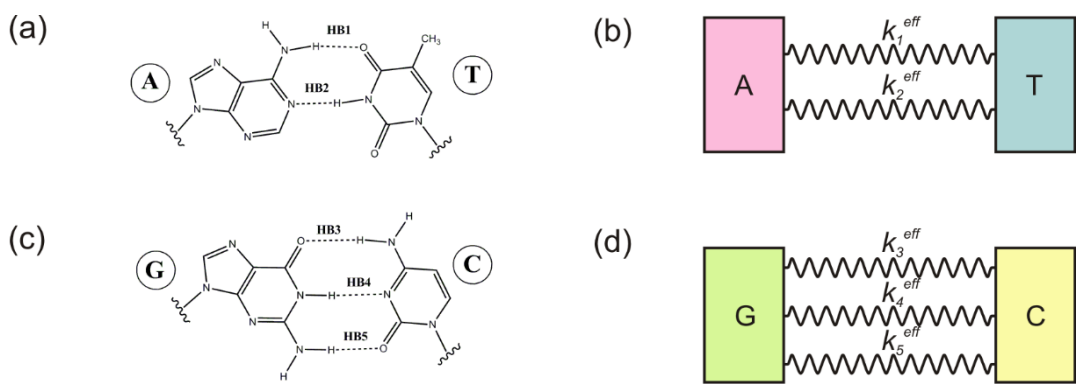

Fig 2. On the electrical nature of a HB. a, Standard dipole model for a HB where two opposite charges represent the interaction. Here and then, H, N and O indicate the position of hydrogen, nitrogen and oxygen atoms respectively. E is the vector representing the electric field in the main axis, Φ the deviation angle, di is the inflection point and d is the axis containing the two atoms involved in the HB. b, Our proposed electrostatic model for a HB interaction, two

Full version is available on: http://journals.plos.org/plosone/article?id=10.1371/journal.pone.0185638

positive charges surrounded by an electron cloud, details in the main text. c, Heaviside representation for $\Phi = \Phi$ (d), the behaviour expected for the model included in b) in the absence of electron cloud. d, Representation of the model included in b) approached as a limiting Heaviside function, see eq. S2 in the S3 File for the definition of finite t=ta values.

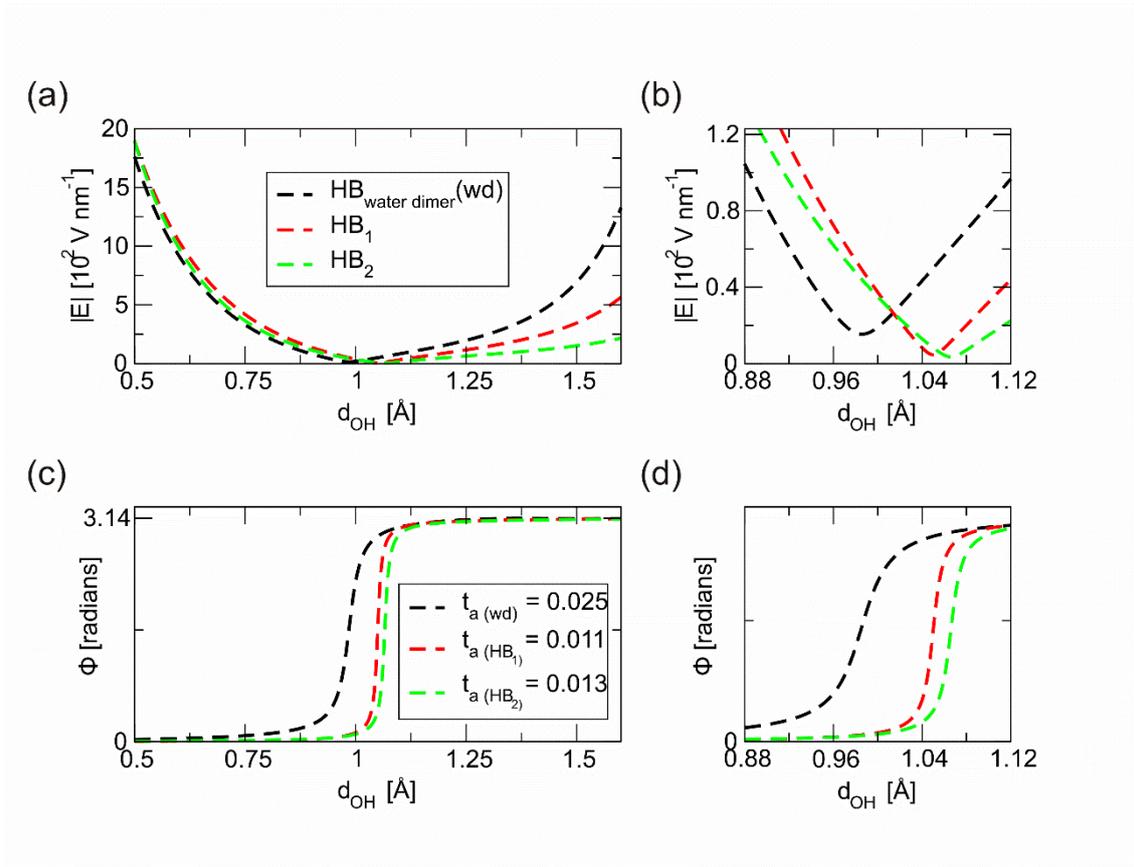

Fig 3. Electrical characterization of inter pairs HB in a model A/T base comparing with the HB stabilizing a water dimer sample. In **a**) and **b**), we represent the variations of the electric field modulus ($|\mathbf{E}|$) along the two-atom axis. The asymptotic behaviour coincides with atomic positions, positive charge centres with the coordinates origin in the atom, O or N, bridging the hydrogen atom. The existence of finite non-zero minimum values ($|\mathbf{E}|_{threshold} \sim 10$ V nm-1) is essential for understanding the electrical inertia as well as the mechanical response originated in DNA. **c**) and **d**), show the characteristic evolution of the angle between **E** relative to the two

atom axis, $\Phi = \Phi(d)$. The deviations of such curves from the ideal Heaviside function are an indication of the electric susceptibility of chemical bonds, more details are included in the main text. The descriptors included in the inserted table accounts for electromechanical information. The numerical values, compared with the HB the water dimer ($t_{a(\text{wd})}$) are indicative of superior strengths in the HB pairing A/T than in the HB forming the water dimer. See S2 File for details on the calculation of the water dimer parameters.

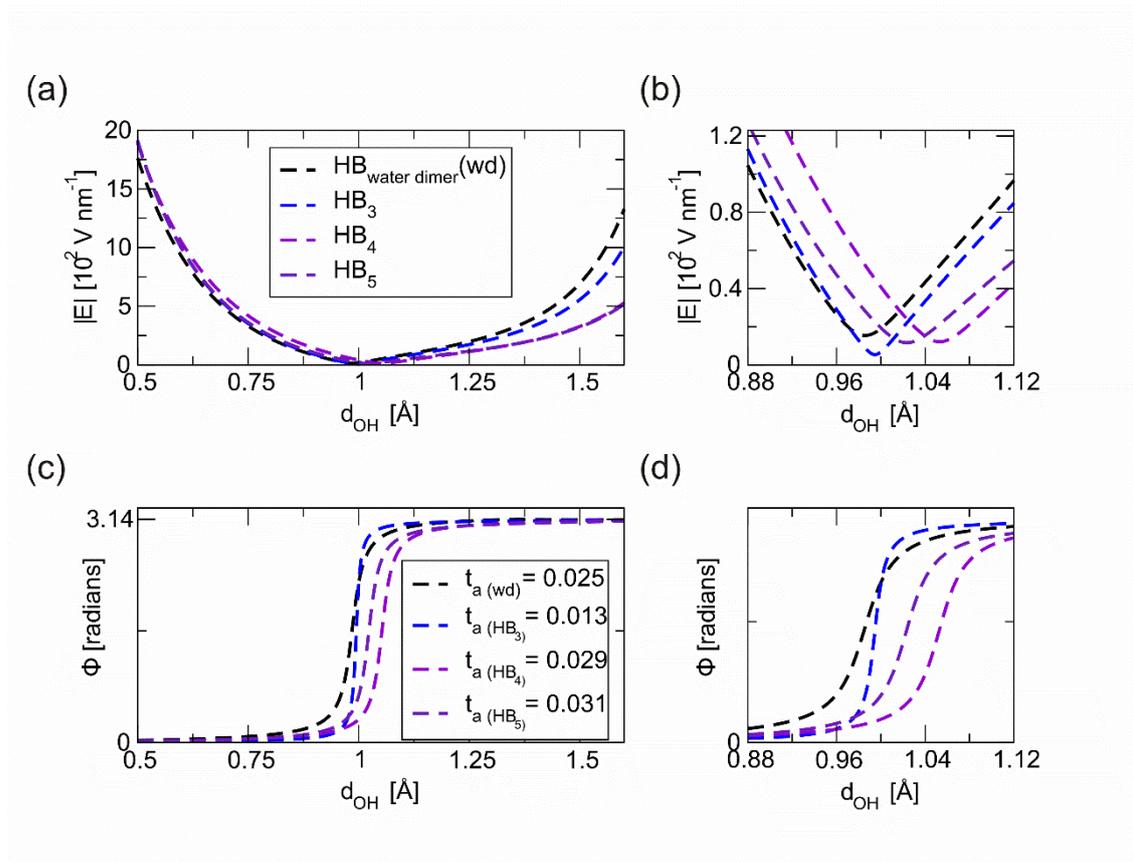

Fig 4. Electrical characterization of inter pairs HB in a model G/C base comparing with the HB stabilizing a water dimer sample. In line with the reasoning included in Fig 3, in **a**) and **b**), we represent the variations of the electric field modulus ($|\mathbf{E}|$) along the two-atom axis. In **c**) and **d**), show the characteristic evolution of the angle between **E** the two atom axis, $\Phi = \Phi(d)$.